\documentclass[twocolumn,aps,pra,showpacs]{revtex4}
\usepackage[T1]{fontenc}
\usepackage[latin9]{inputenc}
\usepackage{amsmath}
\usepackage{amssymb}
\usepackage{graphicx}
\usepackage{esint}

\makeatletter
\@ifundefined{textcolor}{}
{%
 \definecolor{BLACK}{gray}{0}
 \definecolor{WHITE}{gray}{1}
 \definecolor{RED}{rgb}{1,0,0}
 \definecolor{GREEN}{rgb}{0,1,0}
 \definecolor{BLUE}{rgb}{0,0,1}
 \definecolor{CYAN}{cmyk}{1,0,0,0}
 \definecolor{MAGENTA}{cmyk}{0,1,0,0}
 \definecolor{YELLOW}{cmyk}{0,0,1,0}
}

\makeatother

\begin{document}

\title{First and second sound of a unitary Fermi gas in highly oblate harmonic
traps}

\author{Hui Hu$^{1}$, Paul Dyke$^{1}$, Chris J. Vale$^{1}$, and Xia-Ji
Liu$^{1}$}

\email{xiajiliu@swin.edu.au}

\affiliation{$^{1}$Centre for Quantum and Optical Science, Swinburne University
of Technology, Melbourne 3122, Australia}

\date{\today}
\begin{abstract}
We theoretically investigate first and second sound modes of a unitary
Fermi gas trapped in a highly oblate harmonic trap at finite temperatures.
Following the idea by Stringari and co-workers {[}Phys. Rev. Lett.
\textbf{105}, 150402 (2010){]}, we argue that these modes can be described
by the simplified two-dimensional two-fluid hydrodynamic equations.
Two possible schemes - sound wave propagation and breathing mode excitation
- are considered. We calculate the sound wave velocities and discretized
sound mode frequencies, as a function of temperature. We find that
in both schemes, the coupling between first and second sound modes
is large enough to induce significant density fluctuations, suggesting
that second sound can be directly observed by measuring \textit{in-situ}
density profiles. The frequency of the second sound breathing mode
is found to be highly sensitive to the superfluid density. 
\end{abstract}


\maketitle
\tableofcontents{}

\section{Introduction}

Low-energy excitations of a strongly interacting superfluid quantum
system - in which inter-particle collisions are sufficiently strong
to ensure local thermodynamic equilibrium - can be described by the
Landau two-fluid hydrodynamic theory \cite{Tisza1938,Landau1941,KhalatnikovBook,GriffinBook}.
There are two well-known kinds of excitations, referred to as first
and second hydrodynamic modes, which describe the coupled oscillations
of the superfluid and normal fluid components at finite temperatures.
First sound is an ordinary phenomenon in liquid, describing the propagation
of a pressure or density wave that is associated with normal acoustic
sound. The motions of the superfluid and of normal fluid components
are locked in phase. In contrast, second sound is a phenomenon characteristic
of superfluids. It describes the ability to propagate undamped entropy
oscillations, which are essentially opposite phase oscillations of
the two components. In the quantum liquid of superfluid helium, the
velocities of first and second sound were first measured in 1938 by
Findlay \textit{et al}. \cite{Findlay1938} and in 1946 by Peshkov
\cite{Peshkov1946}, respectively.

Ultracold atomic Fermi gases near a Feshbach resonance represent a
new type of strongly interacting quantum system with unprecedented
control over interatomic interactions, dimensionality and purity \cite{Giorgini2008}.
These are anticipated to provide an ideal table-top system to deepen
our understanding of fermionic superfluidity and Landau two-fluid
hydrodynamics. For a Fermi gas at unitarity, first sound modes have
been investigated in great detail, both experimentally and theoretically
\cite{Stringari2004,Kinast2004,Bartenstein2004,Hu2004,Capuzzi2006,Altmeyer2007,Joseph2007,Adhikari2009,Tey2013,Guajardo2013},
particularly near zero temperature. In contrast, second sound mode
is more difficult to address \cite{Taylor2005,He2007,Taylor2008,Taylor2009,Watanabe2010,Hu2010,Bertaina2010,Hou2013,Hou2013b,Hu2013}
and has only recently been observed in experiments \cite{Sidorenkov2013}.
Key to understanding second sound, is knowledge of the superfluid
density \cite{Taylor2008,Salasnich2010,Baym2013}, a quantity which
is still a subject of intense investigation. Theoretically, first
and second sound modes of the Landau two-fluid hydrodynamic equations
were solved by Taylor \textit{et al}. for an isotropically trapped
unitary Fermi gas \cite{Taylor2009}, using an assumed superfluid
density and by developing a variational approach. For general anisotropic
harmonic traps, the Landau hydrodynamic equations are more difficult
to solve. However, for a highly elongated configuration, it was shown
by Bertaina, Pitaevskii and Stringari \cite{Bertaina2010} that, a
simplified one-dimensional (1D) two-fluid hydrodynamic description
may be derived, generalizing the dimensional reduction at zero temperature
\cite{Adhikari2009}. Following this pioneering idea, the propagation
of a second sound wave was strikingly observed in a highly elongated
unitary Fermi gas by the the Innsbruck team \cite{Sidorenkov2013}.
Using the measured second sound velocity data, the simplified 1D hydrodynamic
equations were solved and used to extract the superfluid density \cite{Sidorenkov2013}.
These results show that studies of second sound provide a promising
route towards an accurate determination of the superfluid density.

Here, we would consider a unitary Fermi gas in highly \emph{oblate}
harmonic traps, which can be readily realised in current experiments.
Indeed, such configurations have already been investigated in laboratory
experiments \cite{Martiyanov2010,Frohlich2011,Dyke2011,Orel2011,Sommer2012,Vogt2012}.
The analysis presented here could therefore be directly tested in
future experiments. Our main results are briefly summarized as follows.
Following the ideas of Stringari and co-workers \cite{Bertaina2010,Hou2013b},
we derive the simplified two-dimensional (2D) Landau two-fluid hydrodynamic
equations. Using a variational approach within the local density approximation
\cite{Taylor2005,Taylor2008,Taylor2009}, we fully solve the coupled
2D hydrodynamic equations in the presence of a weak radial harmonic
confinement. The discretized mode frequencies of first and second
sound are calculated. We find that the density fluctuation due to
the second sound mode is significant, revealing that second sound
is directly observable from in-situ density profiles, after an appropriate
modulation of the weak radial trapping potential. The mode frequency
of second sound is found to depend very critically on the form of
the superfluid density. Our quasi-2D analysis is similar to the quasi-1D
study of Stringari and co-workers \cite{Hou2013b}, however, there
are a number of significant differences. Specifically we fully solve
the coupled hydrodynamic equations without assuming that first and
second sound are decoupled. This enables us to calculate the density
fluctuations associated with the discretized second sound modes. From
an experimental point of view, the possibility of measuring discrete
breathing mode frequencies may allow a more accurate determination
of superfluid density.

The remainder of the paper is organized as follows. In the next section,
we outline the reduced 2D universal thermodynamics, satisfied by the
unitary Fermi gas in highly oblate harmonic traps based on the measured
3D equation of state \cite{Ku2012}. In Sec. III, we derive the simplified
2D Landau two-fluid hydrodynamic equations and discuss very briefly
their applicability. In Sec. IV, we consider the propagation of first
and second sound, which may be excited by creating a short perturbation
of the density. The nature of first and second sound excitations in
the quasi-2D unitary Fermi gas is described. In Sec. V, we fully solve
the simplified 2D hydrodynamic equations and calculate the discretized
breathing mode frequencies of first and second sound. The density
fluctuation associated with low-lying second sound modes is discussed
and the dependence of the second sound mode frequency on the form
of superfluid density is examined. Finally, in Sec. V we draw our
conclusions. The appendix presents further details of our numerical
calculations.

\section{2D reduced thermodynamics}

Let us consider a unitary Fermi gas trapped in a highly anisotropic
pancake-like harmonic potential, 
\begin{equation}
V_{ext}\left(r_{\perp},z\right)=\frac{1}{2}m\omega_{\perp}^{2}r_{\perp}^{2}+\frac{1}{2}m\omega_{z}^{2}z^{2},
\end{equation}
with atomic mass $m$ and the radial and axial trapping frequencies
$\omega_{\perp}$, $\omega_{z}$. The aspect ratio of the trap, defined
by $\lambda=\omega_{z}/\omega_{\perp}\gg1$, can be specifically tuned
in experiments. The number of atoms in the typical sound mode experiments
\cite{Bartenstein2004,Kinast2004,Altmeyer2007,Tey2013,Sidorenkov2013}
is about $N\sim10^{5}$. For such a large number of atoms, it is standard
to use the local density approximation \cite{Giorgini2008}, which
amounts to treating the atoms in the position $(r_{\perp},z)$ locally
as a uniform matter, with a local chemical potential given by, 
\begin{equation}
\mu\left(r_{\perp},z\right)=\mu_{0}-V_{ext}\left(r_{\perp},z\right).\label{eq:muloc}
\end{equation}
Here $\mu_{0}$ is the chemical potential at the trap center. The
validity of the local density approximation can be conveniently estimated
by comparing the number of atoms $N\sim10^{5}$ with a threshold $N_{2D}=\lambda\left(\lambda+1\right)$,
below which there is a dimensional crossover from three- to two-dimensions
\cite{Dyke2011,Orel2011}. At Swinburne, we have previously demonstrated
pancake traps with $\lambda\sim60$. Thus, the ratio $N/N_{2D}\sim30$
, indicating that the system could be deeply in the 3D limit where
the local density approximation is well justified.

\subsection{Universal thermodynamics}

A remarkable feature of a uniform unitary Fermi gas is that all its
thermodynamic functions can be expressed as universal functions a
single dimensionless parameter $\mu/(k_{B}T)$ \cite{Ho2004,Hu2007}.
This is simply due to the fact that at unitarity, the \textit{s}-wave
scattering length - the only length scale used to characterize the
short-range interatomic interaction - diverges. Thus, the remaining
length scales are the thermal wavelength 
\begin{equation}
\lambda_{T}\equiv\sqrt{2\pi\hbar^{2}/\left(mk_{B}T\right)}
\end{equation}
and the mean distance between atoms $n^{-1/3}$, where $n$ is the
number density. Accordingly, the energy scales are given by the temperature,
$k_{B}T$, and by the Fermi energy, 
\begin{equation}
k_{B}T_{F}=\frac{\hbar^{2}}{2m}\left(3\pi^{2}n\right)^{2/3},\label{eq:tf3d_homo}
\end{equation}
or, alternatively, by the chemical potential $\mu$. Following the
scaling analysis, all the thermodynamic functions can therefore be
expressed in terms of universal functions that only depends on the
dimensionless parameter $x=\mu/(k_{B}T).$ For the pressure and number
density, these universal functions are given, respectively, by 
\begin{align}
f_{p}^{3D}\left(x\right) & =\left(k_{B}T\right)^{-1}\lambda_{T}^{3}P\left(\mu,T\right),\label{eq:fp3d}\\
f_{n}^{3D}\left(x\right) & =\lambda_{T}^{3}n\left(\mu,T\right).\label{eq:fn3d}
\end{align}
Using the thermodynamic relation $n=(\partial P/\partial\mu)_{T}$,
we find $f_{n}^{3D}(x)=df_{p}^{3D}(x)/dx$. It is easy to see that
knowing $f_{p}^{3D}$ and $f_{n}^{3D}$, we can calculate directly
all the thermodynamic functions of the uniform unitary Fermi gas \cite{Hou2013b}.

\begin{figure}
\begin{centering}
\includegraphics[clip,width=0.48\textwidth]{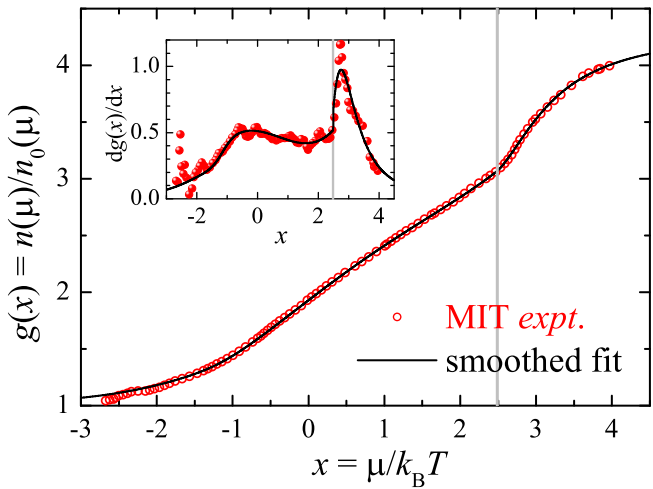} 
\par\end{centering}

\caption{(Color online) The universal scaling function of a strongly interacting
unitary Fermi gas, $n\left(\mu,T\right)/n_{0}(\mu,T)$, where $n_{0}(\mu,T)$
is the number density of an ideal spin-1/2 Fermi gas. The experimental
data from the MIT team (red circles) \cite{Ku2012} are compared with
a smoothed fit (the black line), as described in the text. The inset
shows the comparison for the derivative of the universal scaling function.
The vertical grey lines indicate the critical threshold for superfluidity,
$(\mu/k_{B}T)_{c}\simeq2.49$ \cite{Ku2012}.}

\label{fig1} 
\end{figure}

It is highly non-trivial to theoretically determine the universal
function $f_{p}^{3D}$ or $f_{n}^{3D}$, since there are no small
parameters to control the theory of a strongly interacting Fermi gas
\cite{Liu2005,Hu2006a,Hu2006b,Hu2008,Hu2010b}, except at high temperatures,
where the virial expansion approach is applicable \cite{Liu2009,Liu2013}.
Fortunately, accurate experimental data for the equation of state
are now available from the landmark experiments performed by the Massachusetts
Institute of Technology (MIT) team \cite{Ku2012}. In Fig. \ref{fig1},
we present their data for the universal scaling function, 
\begin{equation}
g\left(x\right)\equiv\frac{n\left(\mu,T\right)}{n_{0}\left(\mu,T\right)}=\frac{f_{n}^{3D}\left(x\right)}{f_{n,0}^{3D}\left(x\right)},
\end{equation}
where the subscript ``0'' indicates the result of an ideal, non-interacting
Fermi gas and 
\begin{equation}
f_{n,0}^{3D}\left(x\right)=\frac{4}{\sqrt{\pi}}\int_{0}^{\infty}dt\frac{\sqrt{t}e^{-t}e^{x}}{1+e^{x}e^{-t}}.
\end{equation}
For later convenience of numerical calculations, we have fitted the
experimental data to an analytic expression, as detailed in the Appendix
A. The fitting curve is smoothly extrapolated to both low and high
temperature regimes where the behavior of $f_{n}^{3D}\left(x\right)$
is well-understood \cite{Hu2010,Hou2013,Liu2009,Liu2013}. As evident
in Fig. \ref{fig1}, the relative error in the fitting is less than
$0.5\%$, significantly smaller than the standard error for the experimental
data (i.e., $2\%$).

\subsection{Local density approximation}

Within the local density approximation, we can write the local pressure
and number density using the universal functions, 
\begin{align}
P\left(r_{\perp},z\right) & =\frac{k_{B}T}{\lambda_{T}^{3}}f_{p}^{3D}\left[\frac{\mu(r_{\perp},z)}{k_{B}T}\right],\\
n\left(r_{\perp},z\right) & =\frac{1}{\lambda_{T}^{3}}f_{n}^{3D}\left[\frac{\mu(r_{\perp},z)}{k_{B}T}\right],
\end{align}
where the local chemical potential is given by Eq. (\ref{eq:muloc}).
For a trapped Fermi gas in three dimensions, the Fermi temperature
$T_{F}$ is defined as \cite{Giorgini2008} 
\begin{equation}
k_{B}T_{F}=\hbar\left(3N\omega_{\perp}^{2}\omega_{z}\right)^{1/3}.
\end{equation}
Using the number equation $N=\int d\mathbf{r}_{\perp}dzn(r_{\perp},z)$,
we relate $\mu_{0}/k_{B}T$ to the reduced temperature $T/T_{F}$
by, 
\begin{equation}
\frac{T}{T_{F}}=\left[\frac{6}{\sqrt{\pi}}\int_{0}^{\infty}dt\sqrt{t}f_{n}^{3D}(\frac{\mu_{0}}{k_{B}T}-t)\right]^{-1/3}.\label{eq:tf3d_trap}
\end{equation}
The onset of superfluidity in a trapped unitary Fermi gas occurs at
$\mu_{0}/k_{B}T=x_{c}\simeq2.49$. Numerically, we find that $T_{c}\simeq0.223T_{F}$,
in agreement with the previous results \cite{Hou2013b,Ku2012}.

\subsection{2D reduced thermodynamic functions}

In highly oblate harmonic traps, although the cloud is still three-dimensional,
its low-energy dynamics are greatly affected by the tight axial confinement.
This situation is very similar to a highly elongated unitary Fermi
gas considered earlier by Bertaina, Pitaevskii and Stringari \cite{Bertaina2010},
who showed that with tight radial confinement, the standard Landau
two-fluid hydrodynamic equations could reduce to a simplified 1D form.
For the same reason, which will be explained in greater detail in
the next section, first and second sound under the tight axial confinement
can be described by a simplified 2D two-fluid hydrodynamic description.
In brief, due to the nonzero viscosity and thermal conductivity, local
fluctuations in temperature ($\delta T$) and chemical potential ($\delta\mu$)
become independent of the axial coordinate, for any low-energy excitations
at the frequency $\omega\sim\omega_{\perp}\ll\omega_{z}$. Therefore,
the axial degree of freedom in all the thermodynamic variables that
enter the Landau two-fluid hydrodynamic equations becomes irrelevant
and can be integrated out. We can then derive 2D reduced thermodynamics
and immediately have the reduced Gibbs-Duhem relation, 
\begin{equation}
\delta P_{2}=s_{2}\delta T+n_{2}\delta\mu,\label{eq:Gibbs_Duhem}
\end{equation}
where the variables $P_{2}$, $s_{2}$ and $n_{2}$ are the axial
integrals of their three-dimensional counterparts, namely the local
pressure, entropy density and number density. The variable $P_{2}$
is given by, 
\begin{equation}
P_{2}\left(r_{\perp}\right)\equiv\int dzP\left(r_{\perp},z\right)=\frac{m\left(k_{B}T\right)^{3}}{\pi^{3/2}\hbar^{3}\omega_{z}}f_{p}\left(x\right),\label{eq:pressure2d}
\end{equation}
where 
\begin{equation}
x\equiv\frac{\left(\mu_{0}-m\omega_{\perp}^{2}r_{\perp}^{2}/2\right)}{k_{B}T}=\frac{\mu\left(r_{\perp}\right)}{k_{B}T}
\end{equation}
and we have introduced the universal scaling function, 
\begin{equation}
f_{p}\left(x\right)\equiv\int_{0}^{\infty}dtf_{p}^{3D}\left(x-t^{2}\right).\label{eq:fp}
\end{equation}
All the 2D thermodynamic variables can be derived from the reduced
Gibbs-Duhem relation, for example, 
\begin{align}
n_{2} & =\left(\frac{\partial P_{2}}{\partial\mu}\right)_{T}=\frac{m\left(k_{B}T\right)^{2}}{\pi^{3/2}\hbar^{3}\omega_{z}}f_{n}\left(x\right),\label{eq:density2d}\\
s_{2} & =\left(\frac{\partial P_{2}}{\partial T}\right)_{\mu}=\frac{m\left(k_{B}T\right)^{2}}{\pi^{3/2}\hbar^{3}\omega_{z}}\left[3f_{p}\left(x\right)-xf_{n}\left(x\right)\right],\label{eq:entropy2d}
\end{align}
where 
\begin{equation}
f_{n}\left(x\right)\equiv\frac{df_{p}\left(x\right)}{dx}=\int_{0}^{\infty}dtf_{n}^{3D}\left(x-t^{2}\right).
\end{equation}
In addition, it is readily shown that the specific heat per particle
at constant column density and pressure are given by \cite{Hou2013},
\begin{align}
\bar{c}_{v2} & =T\left(\frac{\partial\bar{s}_{2}}{\partial T}\right)_{n_{2}}=6\frac{f_{p}\left(x\right)}{f_{n}\left(x\right)}-4\frac{f_{n}\left(x\right)}{f_{n}^{'}\left(x\right)},\label{eq:cv2d}\\
\bar{c}_{p2} & =T\left(\frac{\partial\bar{s}_{2}}{\partial T}\right)_{P_{2}}=\bar{c}_{v2}\left[\frac{3}{2}\frac{f_{p}\left(x\right)f_{n}^{'}\left(x\right)}{f_{n}^{2}\left(x\right)}\right],\label{eq:cp2d}
\end{align}
where $\bar{s}_{2}\equiv s_{2}/(n_{2}k_{B})$ is the entropy per particle
and $f_{n}^{'}(x)\equiv df_{n}(x)/dx$. It is also straightforward
to check the universal relations, 
\begin{align}
\left(\frac{\partial P_{2}}{\partial n_{2}}\right)_{\bar{s}_{2}} & =\frac{3P_{2}}{2n_{2}},\label{eq:dpdn}\\
\left(\frac{\partial P_{2}}{\partial s_{2}}\right)_{n_{2}} & =\frac{T}{2}.\label{eq:dpds}
\end{align}

\begin{figure}
\begin{centering}
\includegraphics[clip,width=0.48\textwidth]{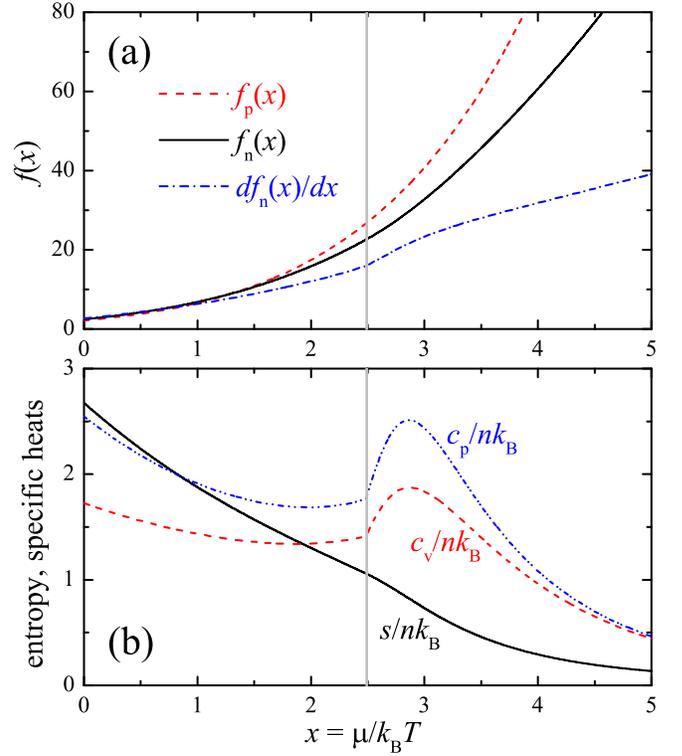} 
\par\end{centering}

\caption{(Color online) (a) 2D universal scaling functions $f_{p}(x)$, $f_{n}(x)$
and $df_{n}(x)/dx$ as a function of the dimensionless variable $x=\mu/(k_{B}T)$.
(b) 2D entropy $\bar{s}=s/(nk_{B})$ and specific heats per particle
$\bar{c}_{v}=c_{v}/(nk_{B})$ and $\bar{c}_{p}=c_{p}/(nk_{B})$ ,
as a function of $x=\mu/(k_{B}T)$. The vertical grey lines indicate
the critical threshold for superfluidity, $x_{c}\simeq2.49$ \cite{Ku2012}.}

\label{fig2} 
\end{figure}

In Fig. \ref{fig2}, we report the relevant 2D universal scaling functions,
calculated by using the experimental MIT data for $f_{n}^{3D}\left(x\right)$
after smoothing.

\subsection{Superfluid density}

We can also express the local superfluid density as a universal (but
as yet undetermined) function $f_{s}^{3D}$: 
\begin{equation}
n_{s}\left(r_{\perp},z\right)=\frac{1}{\lambda_{T}^{3}}f_{s}^{3D}\left[\frac{\mu(r_{\perp},z)}{k_{B}T}\right].
\end{equation}
Integrating out the axial coordinate, we obtain 
\begin{equation}
n_{s2}\left(r_{\perp}\right)=\int dzn_{s}\left(r_{\perp},z\right)=\frac{m\left(k_{B}T\right)^{2}}{\pi^{3/2}\hbar^{3}\omega_{z}}f_{s}\left(x\right),\label{eq:sfdensity2d}
\end{equation}
where the universal scaling function $f_{s}(x)$ is given by, 
\begin{equation}
f_{s}\left(x\right)=\int_{0}^{\infty}dtf_{s}^{3D}\left(x-t^{2}\right).\label{eq:fs}
\end{equation}

\begin{figure}
\begin{centering}
\includegraphics[clip,width=0.48\textwidth]{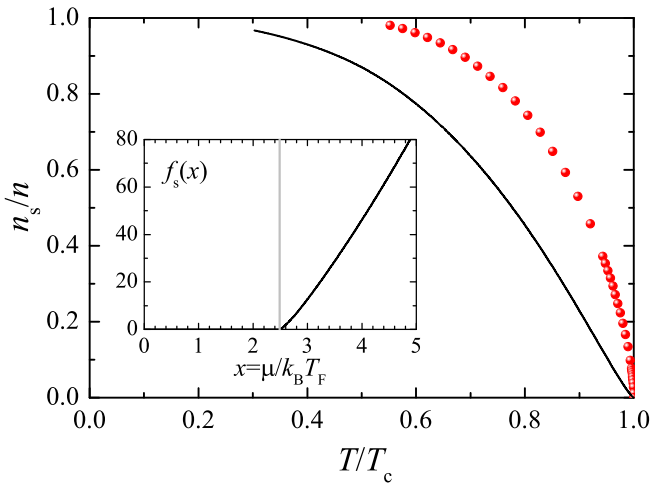} 
\par\end{centering}

\caption{(Color online) Bulk superfluid fraction of superfluid helium (red
circles) \cite{Dash1957} and the corresponding 2D reduced superfluid
fraction (the black line). The inset shows the 2D universal scaling
function for superfluid density, to be used for a unitary Fermi gas.
The vertical grey lines indicate the critical threshold for superfluidity,
$x_{c}\simeq2.49$ \cite{Ku2012}.}

\label{fig3} 
\end{figure}

As mentioned earlier, in contrast to the equation of state, the universal
function for the superfluid density of a unitary Fermi gas $f_{s}^{3D}\left(x\right)$
is not yet known precisely \cite{Taylor2008,Salasnich2010,Baym2013}.
For illustrative purposes in the present work, we will consider the
superfluid fraction of superfluid helium \cite{Dash1957}, 
\begin{equation}
\left(\frac{n_{s}}{n}\right)^{3D}=f_{\textrm{He}}\left(\frac{T}{T_{c}}\right),\label{eq:fhe}
\end{equation}
which is plotted in Fig. \ref{fig3} by red circles. Our choice is
motivated by the similarity for hydrodynamics between superfluid helium
and unitary Fermi gas, presumably arising from strong correlations
in both systems, and justified in part by measurements in Ref. \cite{Sidorenkov2013}.
Using Eq. (\ref{eq:fn3d}), we find that $T/T_{c}=[f_{n}^{3D}(x)/f_{n}^{3D}(x_{c}\simeq2.49)]^{-2/3}$.
Thus, from Eq. (\ref{eq:fhe}), the universal function $f_{s}^{3D}\left(x\right)$
can be calculated using 
\begin{equation}
f_{s}^{3D}\left(x\right)=f_{n}^{3D}\left(x\right)f_{\textrm{He}}\left(\left[\frac{f_{n}^{3D}\left(x\right)}{f_{n}^{3D}\left(x_{c}\simeq2.49\right)}\right]^{-2/3}\right)\label{eq:fs3d}
\end{equation}
and $f_{s}(x)$ is then obtained. In the inset of Fig. 3, we show
$f_{s}(x)$ calculated with the superfluid fraction of superfluid
helium.

In a 2D configuration with $\omega_{\perp}=0$, it is natural to define
a Fermi temperature in terms of the column density $n_{2}$: 
\begin{equation}
k_{B}T_{F}^{2D}=\left(\hbar\omega_{z}\right)^{1/2}\left(\frac{2\pi\hbar^{2}n_{2}}{m}\right)^{1/2},
\end{equation}
which coincides with the usual three-dimensional definition of the
Fermi temperature, Eq. (\ref{eq:tf3d_homo}) . The reduced temperature
is then given by, 
\begin{equation}
\frac{T}{T_{F}^{2D}}=\left[\frac{2}{\sqrt{\pi}}f_{n}\left(x\right)\right]^{-1/2}.
\end{equation}
Therefore, the critical temperature of the unitary Fermi gas in the
pancake geometry is given by, $T_{c}^{2D}=[2f_{n}(x_{c})/\sqrt{\pi}]^{-1/2}T_{F}^{2D}\simeq0.198T_{F}^{2D}$.
In Fig. 3, we show the 2D superfluid fraction $f_{s}/f_{n}$ (the
black line) as a function of $T/T_{c}^{2D}$. It lies systematically
below the three-dimensional superfluid fraction (red circles), due
to the integration over the axial degree of freedom.

It should be noted that the critical behavior of the superfluid density
near the phase transition is greatly affected by the axial integration.
In three dimensions, we may define the critical exponent $\alpha$
by $n_{s}(T\rightarrow T_{c})\sim(T-T_{c})^{\alpha}$ or $f_{s}^{3D}(x\rightarrow x_{c})\sim(x-x_{c})^{\alpha}$.
After integrating out the axial degree of freedom, it is easy to show,
\begin{equation}
f_{s}\left(x\rightarrow x_{c}\right)\sim(x-x_{c})^{\alpha+1/2}.\label{eq:fs_xxc}
\end{equation}
As we shall see, the increase in the critical exponent will reduce
the second sound velocity and hence the second sound mode frequency.

\section{2D simplified two-fluid hydrodynamic equations}

We now derive the simplified 2D Landau two-fluid hydrodynamic equations,
following the idea by Bertaina, Pitaevskii and Stringari \cite{Bertaina2010}.
This 2D picture will generally be valid as we are considering excitations
whose wavelength is long compared to the transverse (axial) cloud
width.

The standard two-fluid equations in the harmonic trap $V_{ext}$ are
given by, 
\begin{align}
m\partial_{t}n+\mathbf{\nabla}\cdot\mathbf{j} & =0,\label{eq:densityHD}\\
m\partial_{t}\mathbf{v}_{s}+\mathbf{\nabla}\left(\mu+V_{ext}\right) & =0,\label{eq:vsHD}\\
\partial_{t}j_{i}+\partial_{i}P+n\partial_{i}V_{ext} & =\partial_{t}\left(\eta\Gamma_{ik}\right),\label{eq:currentHD}\\
\partial_{t}s+\mathbf{\nabla}\cdot\left(s\mathbf{v}_{n}\right) & =\mathbf{\nabla}\cdot\left(\kappa\mathbf{\nabla}T/T\right),\label{eq:entropyHD}
\end{align}
where $\mathbf{j}=m(n_{s}\mathbf{v}_{s}+n_{n}\mathbf{v}_{n})$ is
the current density, $n_{s}$ and $n_{n}$ are the superfluid and
normal density, $\mathbf{v}_{s}$ and $\mathbf{v}_{n}$ are the corresponding
velocity fields, $\Gamma_{ik}\equiv(\partial_{k}v_{ni}+\partial_{i}v_{nk}-2\delta_{ik}\partial_{j}v_{nj}/3)$,
and finally $\eta$ and $\kappa$ are the shear viscosity and thermal
conductivity, respectively. In the above equations, we have kept only
linear terms in the velocity, as we are interested in small-amplitude
dynamics in the linear response regime. Moreover, we have omitted
bulk viscosity terms which give smaller contributions. The Landau
two-fluid hydrodynamic theory is applicable when the mean free path
$l$ is much smaller than the wavelength of the sound wave $\lambda$.
By considering the axial size of the Fermi cloud $R_{z}\sim\sqrt{\hbar/\left(m\omega_{z}\right)}$,
we further require $R_{z}\ll\lambda$, therefore the hydrodynamic
regime in a harmonic trap is achieved when $l\ll R_{z}\ll\lambda$.

As discussed by Stringari and co-workers \cite{Bertaina2010}, in
the presence of tight confinement, the nonzero viscosity and thermal
conductivity may significantly change the low-energy dynamics at the
frequency $\omega\sim\omega_{\perp}\ll\omega_{z}$. This occurs when
the viscous penetration depth $\delta$ - the typical length scale
at which an excitation becomes damped due to shear viscosity - fulfils
the condition, 
\begin{equation}
\delta\equiv\sqrt{\frac{\eta}{mn_{n}\omega}}\gg R_{z}.\label{eq:criterionBPS}
\end{equation}
In the case of superfluid helium in a thin capillary, the above condition
makes the normal component of the liquid stick to the wall and thus
the normal velocity field vanishes \cite{Atkins1959}. With a soft
wall caused by the tight axial confinement, the normal component can
move but the normal velocity field becomes uniform along the axial
direction \cite{Bertaina2010}. For the same reason, the superfluid
velocity also becomes independent of the axial coordinate. Analogously,
a nonzero thermal conductivity in Eq. (\ref{eq:entropyHD}) for the
entropy conservation leads to the independence of the temperature
fluctuation $\delta T$ on the axial coordinate. The tight axial confinement
also implies that the axial component of the velocity fields, for
both superfluid and normal fluid, must be much smaller than the transverse
one and may be neglected. Using Eq. (\ref{eq:vsHD}) for the superfluid
velocity implies that the chemical potential fluctuation $\delta\mu$
is essentially independent of the axial coordinate. In brief, we conclude
that, owing to the crucial roles played by the shear viscosity and
thermal conductivity, under the tight axial confinement the velocity
fields are independent of the axial position and a global thermal
equilibrium along the axial direction is established.

With these observations, it is straightforward to write down the simplified
two-fluid hydrodynamic equations by integrating out the axial degree
of freedom in Eqs. (\ref{eq:densityHD}), (\ref{eq:vsHD}), (\ref{eq:currentHD})
and (\ref{eq:entropyHD}), which take the following forms, 
\begin{align}
m\partial_{t}n_{2}+\mathbf{\nabla_{\perp}}\cdot\mathbf{j}_{\perp} & =0,\label{eq:densityHD2d}\\
m\partial_{t}\mathbf{v}_{s\perp}+\mathbf{\nabla}_{\perp}\left[\mu\left(r_{\perp}\right)+V_{ext}\left(r_{\perp}\right)\right] & =0,\label{eq:vsHD2d}\\
\partial_{t}\mathbf{j}_{\perp}+\nabla_{\perp}P_{2}+n_{2}\nabla_{\perp}V_{ext} & =0,\label{eq:currentHD2d}\\
\partial_{t}s_{2}+\mathbf{\nabla}_{\perp}\cdot\left(s_{2}\mathbf{v}_{n\perp}\right) & =0,\label{eq:entropyHD2d}
\end{align}
where the current density now becomes $\mathbf{j_{\perp}}=m(n_{s2}\mathbf{v}_{s\perp}+n_{n2}\mathbf{v}_{n\perp})$
and we have omitted the residual dissipation terms along the weakly
confined direction, as in a uniform fluid the effect of viscosity
and thermal conductivity is irrelevant in the long-wavelength limit.

We note that, by introducing a characteristic collisional time $\tau$
related to viscosity, $\tau\simeq\eta/(mn\bar{v}^{2})\sim\eta/(n\hbar\omega_{z})$,
where $\bar{v}$ is the average velocity of atoms and is of the order
of the Fermi velocity $v_{F}\sim\sqrt{\hbar\omega_{z}/m}$, it is
easy to check that Eq. (\ref{eq:criterionBPS}) is equivalent to requiring
the low frequency condition, 
\begin{equation}
\omega\ll\omega_{z}^{2}\tau.
\end{equation}
Recalling that $\omega\sim\omega_{\perp}$ and $\omega_{z}=\lambda\omega_{\perp}$,
the above requirement for achieving the reduced 2D hydrodynamics can
be rewritten as $\lambda^{2}\omega\tau\gg1$. For a highly oblate
trap with typical aspect ratio $\lambda\sim60\gg1$, this condition
is compatible with the condition to enter the hydrodynamic regime,
$l\ll\lambda$, which can alternatively be written as $\omega\tau\ll1$.

\subsection{Variational reformulation of the simplified Landau hydrodynamic equations}

A convenient way to solve the simplified two-fluid hydrodynamic equations
is to reformulate them using Hamilton's variational principle \cite{Zilsel1950}.
Following previous work \cite{Taylor2005,Taylor2008}, the hydrodynamic
modes of these equations with frequency $\omega$ at temperature $T$
can be obtained by minimizing a variational action, which, in terms
of displacement fields $u_{s}(\mathbf{r}_{\perp})=i\omega v_{s\perp}(\mathbf{r}_{\perp})$
and $u_{n}(\mathbf{r}_{\perp})=i\omega v_{n\perp}(\mathbf{r}_{\perp})$
\cite{note1}, takes the following form, 
\begin{eqnarray}
S & = & \frac{1}{2}\int d\mathbf{r}_{\perp}\left[m\omega^{2}\left(n_{s2}\mathbf{u}_{s}^{2}+n_{n2}\mathbf{u}_{n}^{2}\right)-\left(\frac{\partial\mu}{\partial n_{2}}\right)_{s_{2}}\left(\delta n\right)^{2}\right.\nonumber \\
 &  & \left.-2\left(\frac{\partial T}{\partial n_{2}}\right)_{s_{2}}\delta n\delta s-\left(\frac{\partial T}{\partial s_{2}}\right)_{n_{2}}\left(\delta s\right)^{2}\right].\label{eq:usunAction}
\end{eqnarray}
Here, $n_{s2}(r_{\perp})$ and $n_{n2}(r_{\perp})=n_{2}(r_{\perp})-n_{s2}(r_{\perp})$
are the 2D reduced superfluid and normal-fluid densities at equilibrium,
as discussed in the previous section Sec. II. The density fluctuation
$\delta n$ and entropy fluctuation $\delta s$ are given by, 
\begin{equation}
\delta n\left(\mathbf{r}_{\perp}\right)\equiv-\mathbf{\nabla}_{\perp}\cdot\left(n_{s2}\mathbf{u}_{s}+n_{n2}\mathbf{u}_{n}\right)
\end{equation}
and 
\begin{equation}
\delta s\left(\mathbf{r}_{\perp}\right)\equiv-\mathbf{\nabla}_{\perp}\cdot\left(s_{2}\mathbf{u}{}_{n}\right),
\end{equation}
respectively. The effect of the weak radial trapping potential $V_{ext}(r_{\perp})=m\omega_{\perp}^{2}r_{\perp}^{2}/2$
enters the action Eq. (\ref{eq:usunAction}) through the coordinate
dependence of the equilibrium thermodynamic variables $(\partial\mu/\partial n_{2})_{s_{2}}$,
$(\partial T/\partial n_{2})_{s_{2}}$ and $(\partial T/\partial s_{2})_{n_{2}}$,
within the local density approximation. We stress that, all these
thermodynamic variables can be obtained by using the reduced Gibbs-Duhem
relation Eq. (\ref{eq:Gibbs_Duhem}) and can be expressed by the universal
functions $f_{p}(x)$ and $f_{n}(x)$.

In superfluid helium, the solutions of the Landau two-fluid hydrodynamic
equations can be well understood as density and entropy (temperature)
waves, which are the pure in-phase mode with $\mathbf{u}_{s}=\mathbf{u}_{n}$
and the pure out-of-phase mode with $n_{s}\mathbf{u}_{s}+n_{n}\mathbf{u}_{n}=0$,
known as first and second sound, respectively \cite{Landau1941,KhalatnikovBook,GriffinBook}.
For a strongly interacting unitary Fermi gas, we could use the same
classification \cite{Taylor2009}. For this purpose, we rewrite the
action Eq. (\ref{eq:usunAction}) in terms of two new displacement
fields 
\begin{equation}
\mathbf{u}_{a}=\left(n_{s2}\mathbf{u}_{s}+n_{n2}\mathbf{u}_{n}\right)/n_{2}
\end{equation}
and 
\begin{equation}
\mathbf{u}_{e}=\mathbf{u}_{s}-\mathbf{u}_{n},
\end{equation}
considering that the density and temperature fluctuations are given
by 
\begin{equation}
\delta n=-\mathbf{\nabla}_{\perp}\cdot\left(n_{2}\mathbf{u}_{a}\right)
\end{equation}
and 
\begin{equation}
\delta T=\left(\frac{\partial T}{\partial s_{2}}\right)_{n_{2}}\mathbf{\nabla}_{\perp}\cdot\left(\frac{s_{2}n_{s2}}{n_{2}}\mathbf{u}_{e}\right),
\end{equation}
respectively. Ideally, first sound is characterized by $\delta n\neq0$
but $\delta T=0$ and second sound by $\delta n=0$ but $\delta T\neq0$.

Using the standard thermodynamic identities derived from the reduced
Gibbs-Duhem relation Eq. (\ref{eq:Gibbs_Duhem}), after some straightforward
but lengthy algebra, we arrive at 
\begin{equation}
S=\frac{1}{2}\int d\mathbf{r}_{\perp}\left[\mathcal{S}^{(a)}+2\mathcal{S}^{(ae)}+\mathcal{S}^{(e)}\right],\label{eq:uaueAction}
\end{equation}
where\begin{widetext} 
\begin{gather}
\mathcal{S}^{(a)}=m\omega^{2}n_{2}\mathbf{u}_{a}^{2}+\left(\mathbf{\nabla}_{\perp}n_{2}\cdot\mathbf{u}_{a}\right)\left(\mathbf{\nabla}_{\perp}V_{ext}\cdot\mathbf{u}_{a}\right)+2\left(n_{2}\mathbf{\nabla}_{\perp}V_{ext}\cdot\mathbf{u}_{a}\right)\left(\mathbf{\nabla}_{\perp}\cdot\mathbf{u}_{a}\right)-n_{2}\left(\frac{\partial P_{2}}{\partial n_{2}}\right)_{\bar{s}_{2}}\left(\mathbf{\nabla_{\perp}}\cdot\mathbf{u}_{a}\right)^{2},\label{eq:S(a)}\\
\mathcal{S}^{(ae)}=\left(\frac{\partial P_{2}}{\partial s_{2}}\right)_{n_{2}}\left(\mathbf{\nabla}_{\perp}\cdot\mathbf{u}_{a}\right)\left[\mathbf{\nabla}_{\perp}\cdot\left(\frac{s_{2}n_{s2}}{n_{2}}\mathbf{u}_{e}\right)\right],\label{eq:S(ae)}\\
\mathcal{S}^{(e)}=m\omega^{2}\frac{n_{s2}n_{n2}}{n_{2}}\mathbf{u}_{e}^{2}-\left(\frac{\partial T}{\partial s_{2}}\right)_{n_{2}}\left[\mathbf{\nabla}_{\perp}\cdot\left(\frac{s_{2}n_{s2}}{n_{2}}\mathbf{u}_{e}\right)\right]^{2}.\label{eq:S(e)}
\end{gather}
\end{widetext}It is clear that the first and second sound are governed
by the actions $\mathcal{S}^{(a)}$ and $\mathcal{S}^{(e)}$, respectively.
The coupling between first and second sound is controlled by the coupling
term $\mathcal{S}^{(ae)}$, which is in general nonzero. Indeed, in
our case, as $(\partial P_{2}/\partial s_{2})_{n_{2}}=T/2$, strictly
speaking the first and second sound are coupled at any finite temperatures.

\section{Free-propagating first and second sound}

\begin{figure}
\begin{centering}
\includegraphics[clip,width=0.48\textwidth]{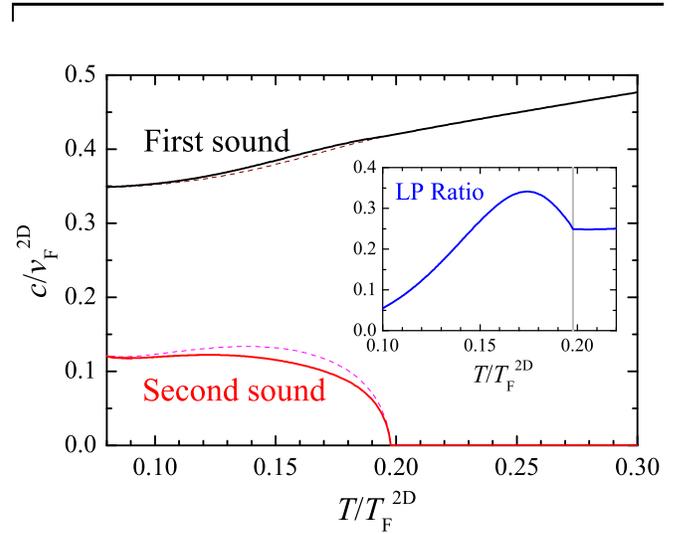} 
\par\end{centering}

\caption{(Color online) 2D first and second sound velocities (solid lines)
as a function of temperature, in units of the Fermi velocity $v_{F}^{2D}=\sqrt{2k_{B}T_{F}^{2D}/m}$
of an ideal Fermi gas at zero temperature at the trap center. The
dashed lines give the decoupled sound velocities. The inset shows
the Landau-Placzek parameter $\epsilon_{\textrm{LP}}$. The vertical
grey lines indicate the critical temperature for superfluidity, $T_{c}\simeq0.198T_{F}^{2D}$,
in the absence of harmonic confinement in the transverse direction.}

\label{fig4} 
\end{figure}

For a uniform superfluid ($V_{ext}=0$), the solutions of $\mathcal{S}^{(a)}$
and $\mathcal{S}^{(e)}$ are plane waves of wave vector $q$ with
dispersion $\omega_{1}=c_{1}q$ and $\omega_{2}=c_{2}q$, where 
\begin{equation}
c_{1}=\sqrt{\frac{1}{m}\left(\frac{\partial P_{2}}{\partial n_{2}}\right)_{\bar{s}_{2}}}\label{eq:c1}
\end{equation}
and 
\begin{equation}
c_{2}=\sqrt{\frac{k_{B}T}{m}\frac{\bar{s}_{2}^{2}}{\bar{c}_{v2}}\frac{n_{s2}}{n_{n2}}}.\label{eq:c2_cv}
\end{equation}
These expressions for first and second sound velocities are the standard
results used to describe superfluid helium, when the corresponding
equation of state and superfluid density are used \cite{Landau1941,KhalatnikovBook,GriffinBook}.
In Fig. \ref{fig4}, we show the decoupled first and second sound
velocities of a highly oblate Fermi gas at unitarity, by using dashed
lines.

Including the coupling term $\mathcal{S}^{(ae)}$ and using the standard
thermodynamic relations, it is straightforward to show that the solutions
for sound velocities $u$ of the simplified two-fluid hydrodynamic
equations satisfy the following equation, 
\begin{equation}
u^{4}-u^{2}\left(c_{1}^{2}+c_{2}^{2}\right)+\frac{c_{1}^{2}c_{2}^{2}}{\gamma}=0,\label{eq:soundvelocity}
\end{equation}
where 
\begin{equation}
\gamma\equiv\frac{\left(\partial P_{2}/\partial n_{2}\right)_{\bar{s}_{2}}}{\left(\partial P_{2}/\partial n_{2}\right)_{T}}=\frac{\bar{c}_{p2}}{\bar{c}_{v2}}>1.
\end{equation}
It gives rise to two solutions for the sound velocity, $u_{1}$ and
$u_{2}$, which in the absence of the coupling term $\mathcal{S}^{(ae)}$
(i.e., $\gamma=1$), coincide with the decoupled first and second
sound velocities, $c_{1}$ and $c_{2}$. The numerical results for
$u_{1}$ and $u_{2}$ are shown in Fig. \ref{fig4} by solid lines.
The temperature dependence of the reduced sound velocities in Fig.
\ref{fig4} is very similar to that of a 3D unitary Fermi gas predicted
in the earlier works \cite{Hu2010,Salasnich2010}. We attribute this
qualitative similarity to the strongly interacting nature of the system.

In the case of a small parameter $\theta\equiv c_{2}^{2}/(\gamma c_{1}^{2})\ll1$,
which is indeed true for a highly oblate unitary Fermi gas, we may
solve Eq. (\ref{eq:soundvelocity}) perturbatively. We find the expansions
\cite{Hu2010}, 
\begin{align}
u_{1}^{2} & =c_{1}^{2}\left[1+\left(\gamma-1\right)\theta+\cdots\right],\\
u_{2}^{2} & =\frac{c_{2}^{2}}{\gamma}\left[1-\left(\gamma-1\right)\theta+\cdots\right].
\end{align}
To the leading order of $\theta$, the first sound velocity is not
affected by the coupling term, but the second sound velocity decreases
by a factor of $\sqrt{\gamma}$ and is now given by \cite{Hu2010,Hou2013b},
\begin{equation}
u_{2}=\sqrt{\frac{k_{B}T}{m}\frac{\bar{s}_{2}^{2}}{\bar{c}_{p2}}\frac{n_{s2}}{n_{n2}}}.\label{eq:c2_cp}
\end{equation}
Therefore, quantitatively, the coupling between first and second sound
can be characterized by the so-called Landau-Placzek (LP) parameter
$\epsilon_{\textrm{LP}}\equiv\gamma-1$ \cite{Hu2010}. In the inset
of Fig. 4, we plot the LP parameter as a function of temperature.
In the superfluid phase, it is always smaller than $0.4$, indicating
that the first and second sound couple very weakly in a unitary Fermi
gas. A similar situation happens in superfluid helium, where $\bar{c}_{p}\simeq\bar{c}_{v}$
and hence $\epsilon_{\textrm{LP}}\simeq0$ \cite{Taylor2008}. From
the above approximate expression for the second sound velocity and
Eq. (\ref{eq:fs_xxc}) for 2D superfluid density, it is clear the
velocity vanishes as 
\begin{equation}
u_{2}\sim\left(1-\frac{T}{T_{c}}\right)^{\alpha/2+1/4},\label{eq:u2_ttc}
\end{equation}
when approaching to the superfluid phase transition from below. Here,
$\alpha\simeq2/3$ is the critical exponent of the superfluid density
of a unitary Fermi gas in three dimensions.

\begin{figure}
\begin{centering}
\includegraphics[clip,width=0.48\textwidth]{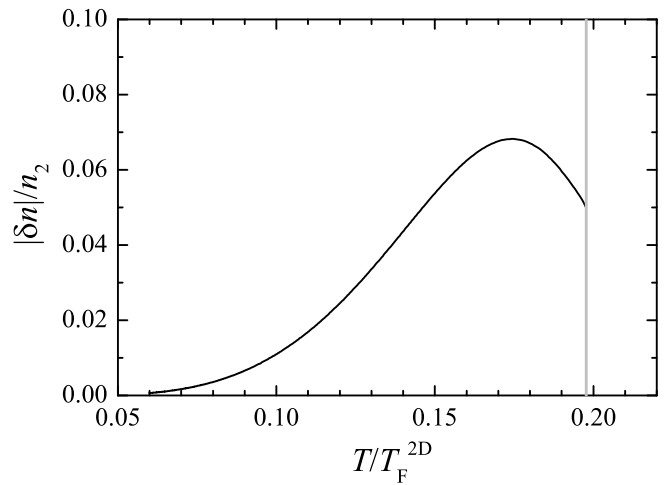} 
\par\end{centering}

\caption{(Color online) The density fluctuation of 2D second sound, calculated
in the case of $\delta T/T=10\%$. The vertical grey lines indicate
the critical temperature for superfluidity, $T_{c}\simeq0.198T_{F}^{2D}$,
in the absence of harmonic confinement in the transverse direction.}

\label{fig5} 
\end{figure}

Although the coupling between first and second sound is weak in a
unitary Fermi gas, it is of significant importance for the purpose
of experimental detection of second sound. Unlike superfluid helium,
temperature oscillations - the characteristic motion of the second
sound - are difficult to observe in ultracold atomic gases. Density
measurement is the most efficient way to characterize any low-energy
dynamics. As a result, an observable second sound mode must have a
sizable density fluctuation, which can only be induced by its coupling
to first sound modes. The strength of density fluctuations in second
sound may be conveniently characterized by the following ratio between
the relative density and temperature fluctuations, 
\begin{equation}
\left(\frac{\delta n/n_{2}}{\delta T/T}\right)_{2\textrm{nd}}\simeq\frac{T}{n_{2}}\left(\frac{\partial n_{2}}{\partial T}\right)_{P_{2}}=2-3\frac{f_{p}f_{n}^{'}}{f_{n}^{2}},\label{eq:ratio2nd}
\end{equation}
where the subscript ``2nd'' indicates the second sound and we have
taken the derivative at const pressure, as the second sound is properly
considered as an oscillating wave at constant pressure, rather than
at constant density, as indicated by Eq. (\ref{eq:c2_cp}) \cite{Hou2013b}.
In contrast, the ratio between the relative density and temperature
fluctuations in first sound is given by,
\begin{equation}
\left(\frac{\delta n/n_{2}}{\delta T/T}\right)_{1\textrm{st}}\simeq\frac{T}{n_{2}}\left(\frac{\partial n_{2}}{\partial T}\right)_{\bar{s}_{2}}=2.\label{eq:ratio1st}
\end{equation}

In Fig. \ref{fig5}, we report the density fluctuation of second sound
as a function of temperature in the superfluid phase, calculated with
an assumed temperature fluctuation ratio $\delta T/T=10\%$. This
is a typical temperature fluctuation, achievable in current experiments.
For example, in the recent first-sound collective mode measurement,
it was shown that $(\delta n/n_{2})_{1\textrm{st}}\sim20\%$ over
a wide temperature window \cite{Tey2013,Guajardo2013}. By using Eq.
(\ref{eq:ratio1st}), we therefore assume that a temperature fluctuation
at $\delta T/T=10\%$ can be easily excited. From Fig. \ref{fig5},
it is easy to see that the density fluctuation of second sound is
very significant when temperature $T>0.7T_{c}\sim0.14T_{F}^{2D}$,
revealing that the propagation of a second sound pulse could be experimentally
detected via density measurements. This result suggests that a similar
approach to that demonstrated by the Innsbruck experiment \cite{Sidorenkov2013}
could be applied in the oblate geometry considered here, where a focused
blue-detuned laser propagating along the tightly confined direction
could excite a second sound wave which will propagate radially outwards
from the cloud centre. Subsequent absorption images taken at different
times after the laser pulse could record the density fluctuation induced
by the propagating second sound wave.

While the essential physics of our scheme is the same as that considered
by by Stringari and co-workers in highly elongated trap \cite{Hou2013b},
the quasi-2D geometry may offer advantages for experiments. Specifically,
clouds confined in elongated traps typically have very high peak densities
leading to high optical densities that are difficult to measure quantitatively
using absorption imaging \cite{Reinaudi2007,Esteve2008}. This means
quantifying small density fluctuations in such clouds can prove challenging.
In a quasi-2D trap however, such high peak optical densities can be
avoided and images taken along the direction of tight confinement
can be integrated over the angular coordinate to obtain an accurate
measure of the radial density. Small density fluctuations should therefore
be more easily detected.

\section{Breathing first and second sound modes in harmonic traps}

We now consider a weak transverse harmonic trap, $V_{ext}(r_{\perp})=m\omega_{\perp}^{2}r_{\perp}^{2}/2$
and fully solve the simplified Landau two-fluid hydrodynamic equations
using a variational approach, developed in the previous work \cite{Taylor2008,Taylor2009}.
We focus on compressional (breathing) modes with the projected angular
momentum $l_{z}=0$, since these modes are the easiest one to excite
experimentally.

\subsection{Variational approach}

For breathing modes, we assume the following polynomial ansatz for
the displacement fields: 
\begin{align}
\mathbf{u}_{a}(r_{\perp}) & =\mathbf{\hat{r}}_{\perp}\sum_{i=0}^{N_{p}-1}A_{i}\tilde{r}_{\perp}^{i+1},\\
\mathbf{u}_{e}\left(r_{\perp}\right) & =\mathbf{\hat{r}}_{\perp}\sum_{i=0}^{N_{p}-1}B_{i}\tilde{r}_{\perp}^{i+1},
\end{align}
where $\mathbf{\hat{r}}_{\perp}$ is the unit vector in the radial
direction, $\{A_{i},B_{i}\}$ ($i=0,\cdots,N_{p}-1$) are the $2N_{p}$
variational parameters, and $\tilde{r}_{\perp}\equiv r_{\perp}/R_{F}$
is the dimensionless radial coordinate with $R_{F}$ being the Thomas-Fermi
radius. By inserting this variational ansatz into the action Eq. (\ref{eq:uaueAction}),
we express the action $S$ as functions of the $2N_{p}$ variational
parameters $\{A_{i},B_{i}\}$ . The mode frequencies are obtained
by minimizing the action $S$ with respect to these $2N_{p}$ parameters.
The accuracy of our variational calculations can be improved by increasing
the value of $N_{p}$. It turns out that the calculations converge
quickly as $N_{p}$ increases.

In more detail, it is straightforward to show that, the expression
of the action can be written in a compact form, 
\begin{equation}
S=\frac{1}{2}\Lambda^{\dagger}\mathcal{S}\left(\omega\right)\Lambda,
\end{equation}
where $\Lambda\equiv\left[A_{0},B_{0},\cdots,A_{i},B_{i},\cdots,A_{N_{p}-1},B_{N_{p}-1}\right]^{T}$
and $\mathcal{S}(\omega)$ is a $2N_{p}\times2N_{p}$ matrix with
block elements ($i,j=0,\cdots,N_{p}-1$), 
\begin{equation}
\left[\mathcal{S}\left(\omega\right)\right]_{ij}\equiv\left[\begin{array}{cc}
M_{ij}^{(a)}\omega^{2}-K_{ij}^{(a)} & -K_{ij}^{(ae)}\\
-K_{ji}^{(ae)} & M_{ij}^{(e)}\omega^{2}-K_{ij}^{(e)}
\end{array}\right].\label{eq:matrixsw}
\end{equation}
In $[\mathcal{S}\left(\omega\right)]_{ij}$, we have introduced the
weighted mass moments, 
\begin{align}
M_{ij}^{(a)} & =m\int d\mathbf{r}_{\perp}\tilde{r}_{\perp}^{i+j+2}n_{2}\left(r_{\perp}\right),\\
M_{ij}^{(e)} & =m\int d\mathbf{r}_{\perp}\tilde{r}_{\perp}^{i+j+2}\left[\frac{n_{s2}n_{n2}}{n_{2}}\right]\left(r_{\perp}\right),
\end{align}
and the spring constants, 
\begin{eqnarray}
K_{ij}^{(a)} & = & \frac{3}{2}\frac{\left(i+2\right)\left(j+2\right)}{R_{F}^{2}}\int d\mathbf{r}_{\perp}\tilde{r}_{\perp}^{i+j}P_{2}\left(r_{\perp}\right),\\
K_{ij}^{(ae)} & = & \frac{T}{2}\frac{i\left(i+2\right)}{R_{F}^{2}}\int d\mathbf{r}_{\perp}\tilde{r}_{\perp}^{i+j}\left[\frac{s_{2}n_{s2}}{n_{2}}\right]\left(r_{\perp}\right),\\
K_{ij}^{(e)} & = & \int d\mathbf{r}_{\perp}\left(\frac{\partial T}{\partial s_{2}}\right)_{n_{2}}\frac{1}{r_{\perp}^{2}}\frac{d\left[s_{2}n_{s2}\tilde{r}_{\perp}^{i+2}/n_{2}\right]}{d\tilde{r}_{\perp}}\nonumber \\
 &  & \times\frac{d\left[s_{2}n_{s2}\tilde{r}_{\perp}^{j+2}/n_{2}\right]}{d\tilde{r}_{\perp}}.
\end{eqnarray}
To derive the above equations, we have used the universal relations
satisfied by the highly oblate unitary Fermi gas: $n_{2}(\partial P_{2}/\partial n_{2})_{\bar{s}_{2}}=3P_{2}/2$
and $(\partial P_{2}/\partial s_{2})_{n_{2}}=T/2$. Moreover, we have
used integration by parts to simplify the expressions: for example,
the contributions from the middle two terms in Eq. (\ref{eq:S(a)})
with $V_{ext}(r_{\perp})$ can be shown to cancel with each other
with our polynomial ansatz. For a given value of $\mu_{0}/k_{B}T$
(or $T/T_{F}$, see Eq.(\ref{eq:tf3d_trap})), the weighted mass moments
and spring constants can be calculated by using local thermodynamic
variables in Eqs. (\ref{eq:pressure2d}), (\ref{eq:density2d}), (\ref{eq:entropy2d}),
(\ref{eq:cv2d}) and (\ref{eq:sfdensity2d}). The detailed expressions
for numerical calculations are listed in Appendix B.

It is easy to see that the minimization of the action $S$ is equivalent
to solving 
\begin{equation}
\mathcal{S}\left(\omega\right)\Lambda=0,
\end{equation}
or $\det\mathcal{S}(\omega)=0$. The detailed numerical procedure
is presented in Appendix C. Once a solution (i.e., the mode frequency
$\omega$ and the coefficient vector $\Lambda$) is found, we calculate
the density fluctuation of the mode, using 
\begin{equation}
\delta n\left(r_{\perp}\right)=-\sum_{i=0}^{N_{p}-1}A_{i}\frac{1}{r_{\perp}}\frac{d}{dr_{\perp}}\left[r_{\perp}n_{2}\left(r_{\perp}\right)\tilde{r}_{\perp}^{i+1}\right].
\end{equation}

We have performed numerical calculations for the number of the variational
parameter $N_{p}$ up to $10$, for any given chemical potential $\mu_{0}/k_{B}T$
or temperature $T/T_{F}$. In the following, we first discuss the
decoupled first and second sound. Then, we focus on the effect of
the coupling between first and second sound and the density fluctuation
of second sound modes.

\subsection{Decoupled first and second sound}

\begin{figure}
\begin{centering}
\includegraphics[clip,width=0.48\textwidth]{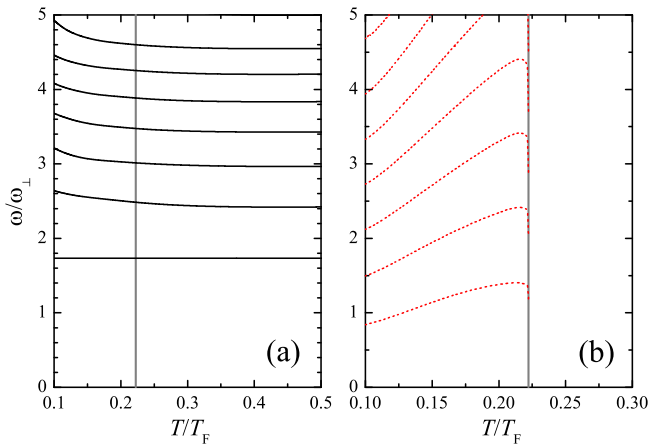} 
\par\end{centering}

\caption{(Color online) The mode frequencies of decoupled first (a) and second
sound (b). The vertical grey lines show the critical temperature of
a three-dimensional trapped unitary Fermi gas, $T_{c}\simeq0.223T_{F}$.}

\label{fig6} 
\end{figure}

In Figs. \ref{fig6}(a) and \ref{fig6}(b), we present mode frequencies
of decoupled first and second sound modes of a highly oblate unitary
Fermi gas, obtained by solving the individual action $\mathcal{S}^{(a)}$
and $\mathcal{S}^{(e)}$, respectively.

For all the first sound modes, except the lowest-lying one, the mode
frequency decreases monotonically with increasing temperature, exhibiting
the same temperature dependence as the first sound modes in an isotropic
\cite{Taylor2009} or highly elongated harmonic trap \cite{Tey2013,Hou2013b}.
The lowest-lying breathing mode instead does not depend on the temperature
and takes an invariant mode frequency $\omega_{B}=\sqrt{3}\omega_{\perp}$.
We note that such a temperature independence is a peculiarity of the
unitary Fermi gas due to its inherent scale invariance. Indeed, our
variational ansatz $\mathbf{u}_{a}(r_{\perp})=\mathbf{r}_{\perp}$
of the lowest-lying breathing mode is one of the few exact scaling
solutions exhibited by the Landau two-fluid hydrodynamic equations
at unitarity \cite{Hou2013}. To understand this, we simply recall
that the spring constant $K_{ij}^{(a)}$ satisfies, 
\begin{equation}
\frac{K_{ij}^{\left(a\right)}}{M_{ij}^{\left(a\right)}}=\frac{3}{2}\frac{\left(i+2\right)\left(j+2\right)}{\left(i+j+2\right)}\omega_{\perp}^{2}.
\end{equation}
Thus, if $i=0$ or $j=0$, we have $K_{ij}^{(a)}=3\omega_{\perp}^{2}M_{ij}^{(a)}$.
Taking into account the fact that $K_{i=0,j}^{(ae)}=0$ or $K_{j=0,i}^{(ae)}=0$,
it is readily seen from Eq. (\ref{eq:matrixsw}) that $\omega^{2}=3\omega_{\perp}^{2}$
provides an exact solution, regardless the number of the variational
ansatz used. We note also that, the first sound solutions converge
very quickly with the number of the variational parameters, indicating
that these solutions are indeed well-approximated by the polynomial
function.

For second sound modes, we find that the mode frequency initially
increases with increasing temperature and then drops to zero very
dramatically when the temperature approaches to the critical value.
This behavior differs from the earlier result in a three-dimensional
isotropic harmonic trap \cite{Taylor2009} but agrees with a recent
prediction made for a high elongated configuration \cite{Hou2013b}.
Qualitatively, we may estimate the discretized second sound mode frequency
by using the expression $\omega\sim u_{2}q$ with a characteristic
wavevector $q\sim1/R_{s}$, where $R_{s}$ is the size of the superfluid
component along the radial direction. Within the local density approximation,
we have $R_{s}\sim R_{F}\sqrt{1-T/T_{c}}$. Using Eq. (\ref{eq:u2_ttc})
$u_{2}\sim(1-T/T_{c})^{\alpha/2+1/4}$, we obtain that 
\begin{equation}
\omega\sim\left(1-\frac{T}{T_{c}}\right)^{\alpha/2-1/4}.
\end{equation}
Thus, for the critical exponent $\alpha>1/2$, the second sound mode
frequency $\omega$ vanishes at $T_{c}$. The superfluid density data
(i.e., of superfluid helium) that we use have a critical exponent
$\alpha\simeq2/3$, and consequently, we find the vanishing frequency
at the transition.

\subsection{Full solutions of 2D two-fluid hydrodynamics}

\begin{figure}
\begin{centering}
\includegraphics[clip,width=0.48\textwidth]{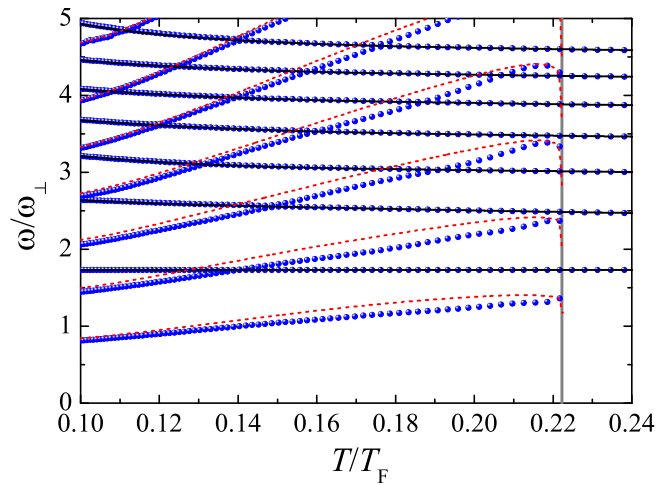} 
\par\end{centering}

\caption{(Color online) Temperature dependence of the full two-fluid hydrodynamic
mode frequencies (blue circles). For comparison, we show also the
mode frequencies of decoupled first and second sound, respectively,
by black solid and red dashed lines. The vertical grey lines show
the critical temperature of a three-dimensional trapped unitary Fermi
gas, $T_{c}\simeq0.223T_{F}$.}

\label{fig7} 
\end{figure}

\begin{figure}
\begin{centering}
\includegraphics[clip,width=0.48\textwidth]{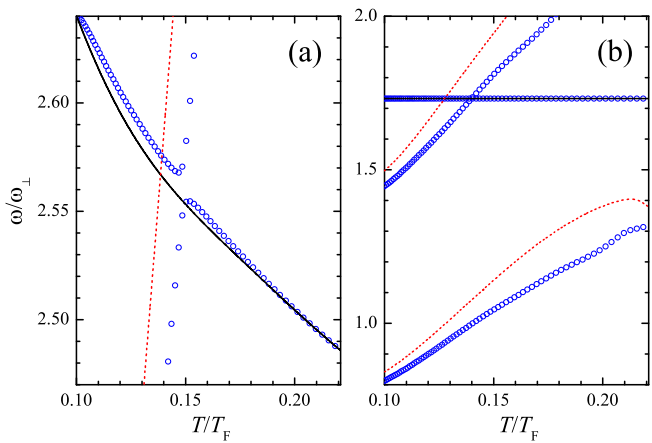} 
\par\end{centering}

\caption{(Color online) Enlarged view of the full two-fluid hydrodynamic mode
frequencies (blue circles), near the $n=2$ first sound mode (a) and
the lowest two second sound modes (b). In the left panel, the avoided
crossing between first and second sound is evident.}

\label{fig8} 
\end{figure}

We now include the coupling term $\mathcal{S}^{(ae)}$. In Fig. \ref{fig7},
we report the full variational results by blue circles. For comparison,
the decoupled first and second sound mode frequencies are also shown,
by black lines and red dashed lines, respectively. As anticipated,
the exact solution with frequency $\omega=\sqrt{3}\omega_{\perp}$
remains unchanged with the inclusion of the coupling term.

Very similar to the sound velocities in the uniform case (see Fig.
\ref{fig4}), the first sound mode frequency is barely affected by
the coupling term $\mathcal{S}^{(ae)}$. This is particularly evident
in Fig. \ref{fig8}(a), in which we show an enlarged view for the
$n=2$ first sound mode (the integer $n$ labels the $n$-th first
sound modes). The correction to the first sound mode frequency due
to the coupling is about $0.5\%$ and occurs only at around $T\sim0.14T_{F}$.

On the other hand, the frequency of second sound modes is notably
pushed down by the coupling term $\mathcal{S}^{(ae)}$, as can be
seen from Fig. \ref{fig8}(b). The maximum correction is up to $20\%$
when the temperature is about $0.18T_{F}$. In the uniform case, i.e.,
Fig. \ref{fig4}, we find a similar correction to the second sound
velocity in the same temperature regime.

\subsection{Density fluctuations of sound modes}

\begin{figure}
\begin{centering}
\includegraphics[clip,width=0.48\textwidth]{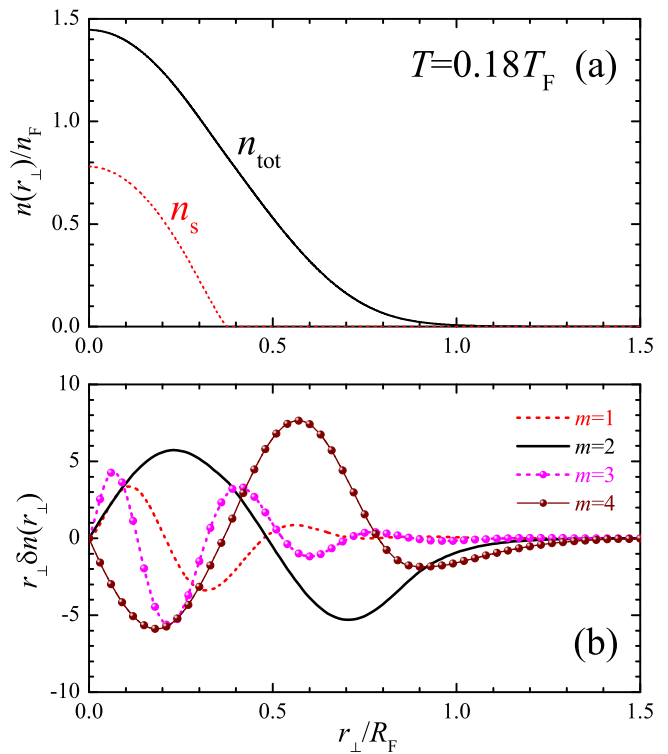} 
\par\end{centering}

\caption{(Color online) (a) Density distribution (the black line) and superfluid
density distribution (the red dashed line) at $T=0.18T_{F}$, in units
of the peak linear density of an ideal Fermi gas at zero temperature
and at the trap center ($n_{F}$). (b) Density fluctuations (in arbitrary
unit) of the lowest two first-sound (solid lines) and second-sound
modes (dashed lines).}

\label{fig9} 
\end{figure}

The sizable correction in the mode frequency of a second sound mode
strongly indicates that, its density fluctuation, as a result of the
coupling to first sound modes, should also be significant. In Fig.
\ref{fig9}(b), we show the density fluctuations of the lowest two
first sound modes ($m=2$ and $4$) and of the lowest two second sound
modes ($m=1$ and $3$), at a temperature $T=0.18T_{F}$. Here, we
have used the integer $m$ as the index of different modes when the
coupling term is taken into account.

Remarkably, the amplitude of the second sound density fluctuations
is just a bit smaller than that of the first sound modes, over a wide
range of temperatures, revealing that it could be detected experimentally.
In this respect, we note that, most recently the density fluctuations
of the low-lying first sound modes have been measured in a highly
elongated unitary Fermi gas, to a reasonably good precision \cite{Tey2013}.
Therefore, it is very likely that a low-lying second sound mode could
also be observed by looking at its density fluctuation, after proper
excitation.

Comparing the density and superfluid density profiles, shown in Fig.
\ref{fig9}(a), we find that the density fluctuation of second sounds
is most significant within the superfluid core, in accordance with
their temperature-wave nature. In contrast, the density fluctuation
of first sound modes extends over the whole Fermi cloud.

\subsection{Dependence on the superfluid density}

\begin{figure}
\begin{centering}
\includegraphics[clip,width=0.48\textwidth]{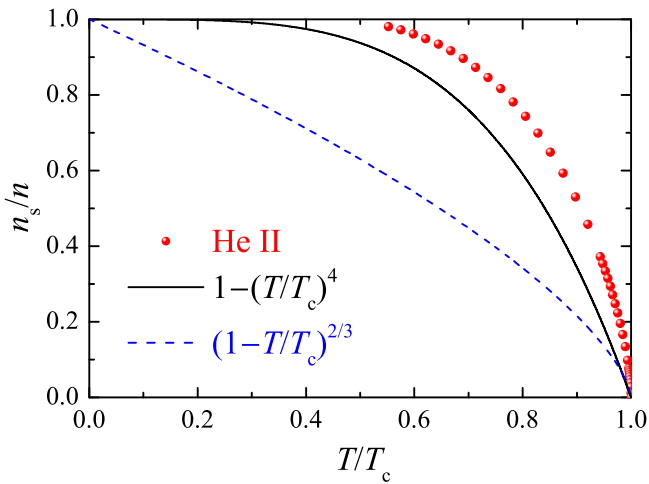} 
\par\end{centering}

\caption{(Color online) Bulk superfluid fraction: the black line and blue dashed
line correspond to the choices, $n_{s}/n=1-(T/T_{c})^{4}$ and $n_{s}/n=(1-T/T_{c})^{2/3}$,
respectively. The red circles are the superfluid fraction of superfluid
helium \cite{Dash1957}. Note that, recently the superfluid fraction
of a 3D unitary Fermi gas has been calculated by Salasnich based on
the Landau's expression for superfluid density and physical elementary
excitations \cite{Salasnich2010}. The predicted temperature dependence
is similar to what we have shown in the figure, presumably due to
the simialr elementary excitations arising from strong interactions.}

\label{fig10} 
\end{figure}

\begin{figure*}
\begin{centering}
\includegraphics[clip,width=0.96\textwidth]{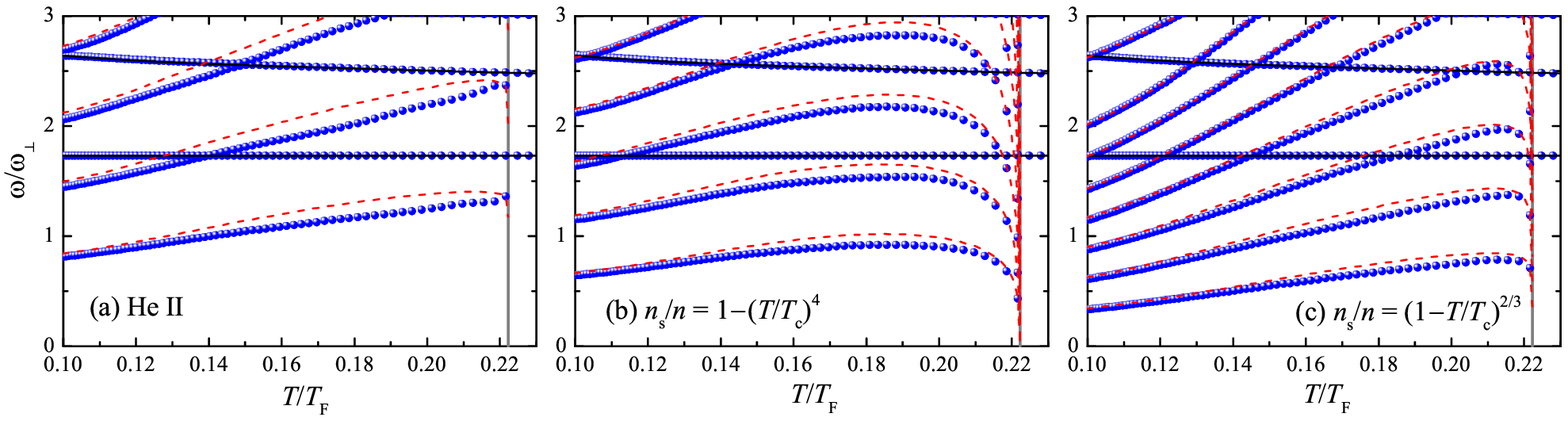} 
\par\end{centering}

\caption{(Color online) Sensitivity of second sound modes on the bulk superfluid
density. The vertical grey lines show the critical temperature of
a three-dimensional trapped unitary Fermi gas, $T_{c}\simeq0.223T_{F}$.}

\label{fig11} 
\end{figure*}

We not that, a complete theoretical description of the superfluid
density for a Fermi gas at unitarity is yet to be determined due to
the theoretical difficulties in handling the strong interaction. In
the above studies, we have considered the superfluid fraction of superfluid
helium for the superfluid density of the unitary Fermi gas, see, for
example, Eqs. (\ref{eq:fhe}) and (\ref{eq:fs3d}). As the second
sound is arguably the most significant demonstration of superfluid,
it is natural to anticipate that the mode frequency of second sound
modes should depend very sensitively on the form of superfluid density.

We have performed numerical calculations with other models of the
superfluid density, as shown in Fig. \ref{fig10}. The two additional
models we considered are: firstly, $(n_{s}/n)^{3D}=1-(T/T_{c})^{4}$,
which is in magnitude similar to the superfluid fraction of superfluid
helium but with a mean-field-like critical exponent $\alpha=1$; secondly,
$(n_{s}/n)^{3D}=(1-T/T_{c})^{2/3}$, which differs significantly from
the superfluid fraction of superfluid helium but takes into account
the correct critical exponent $\alpha\simeq2/3$. The predicted mode
frequencies for different superfluid density are shown in Fig. \ref{fig11}.

It is readily seen that the mode frequency of second sound modes displays
a very sensitive dependence on both the magnitude and critical exponent
of the superfluid density. In particular, the mode frequency with
the superfluid-helium-like superfluid fraction (Fig. \ref{fig11}(a))
differs significantly from that with the choice $(n_{s}/n)^{3D}=1-(T/T_{c})^{4}$,
although these two superfluid fractions differ slightly in magnitude.
The strong dependence is very encouraging, indicating that practically,
the superfluid density of a unitary Fermi gas could be accurately
determined by measuring the mode frequency of low-lying second sound
modes.

\subsection{Experimental considerations}

Ideally, a compressional second sound mode would be excited by modulating
the trapping frequency close to the predicted mode frequency. However,
in our quasi-2D configuration, this can be problematic due to the
fact that the radial confinement is generated by a magnetic field,
whose modulation can also change the \textit{s}-wave scattering length
and hence the system may be away from the unitary regime. We note
that this problem does not exist in other setups where it is possible
to have optical radial (and axial) confinement.

For our experimental setup at Swinburne, therefore, a more practical
scheme would be to locally perturb the cloud and allow it to relax.
Taking inspiration from the selective excitation schemes, developed
by the Innsbruck experiments \cite{Guajardo2013}, one may direct
an appropriately sized blue-detuned laser beam into the cloud and
apply a modulation burst, chosen to provide the best mode matching
with the calculated second sound mode. The power, duration and shape
of the laser beam should be optimized in order to resonantly derive
the desired small amplitude oscillation in the linear response regime.

\section{Conclusions}

In conclusion, we have derived the two-dimensional simplified Landau
two-fluid hydrodynamic equations to describe the low-energy dynamics
of a unitary Fermi gas confined in highly oblate harmonic traps. By
using a variational approach, these two-fluid hydrodynamic equations
have been fully solved. We have discussed in detail the resulting
density-wave (firsts sound) and temperature-wave (second sound) oscillations,
in the absence or presence of a weak transverse harmonic trap. First
and second sound velocities or discretized mode frequencies have been
predicted, accordingly.

We have found a weak coupling between first and second sound in highly
oblate unitary Fermi gas, very similar to the case of superfluid helium.
Though the coupling is weak, it induces significant density fluctuations
for second sound modes, indicating that second sound could be potentially
observed in the highly oblate configuration, by measuring the density
fluctuations. Owing to the strong sensitivity of second sound mode
frequencies to superfluid density, the experimental measurement of
discretized second sound modes could provide a promising way of accurately
determining the superfluid density of a unitary Fermi gas.
\begin{acknowledgments}
We thank Martin W. Zwierlein and Mark J.-H. Ku for providing us their
experimental data. This research was supported by the ARC Discovery
Projects Grant Nos. FT130100815 (HH), DP140103231 (HH), DE140100647
(PD), FT120100034(CJV), DP130101807 (CJV) and DP140100637 (XJL), and
NFRP-China Grant No. 2011CB921502 (XJL, HH).
\end{acknowledgments}
\appendix

\section{Equation of state of a unitary Fermi gas}

In the low temperature superfluid phase, we fit the experimental data
of $g(x)$ by using the Padé approximant of order {[}2/2{]}, 
\begin{equation}
g_{s}\left(x\right)\simeq\xi^{-3/2}\left[\frac{1+p_{1}x^{-1}+p_{2}x^{-2}}{1+q_{1}x^{-1}+q_{2}x^{-2}}\right],
\end{equation}
where $\xi\simeq0.376\pm0.004$ is the Bertsch parameter \cite{Ku2012},
and obtain $p_{1}=-4.49538$, $p_{2}=+5.46140$, $q_{1}=-4.43558$
and $q_{2}=+5.50598$.

For the normal state, i.e., $x\subset[-1.2,x_{c}\simeq2.49]$, we
fit the data with a fourth order polynomial 
\begin{equation}
g_{1}\left(x\right)\simeq a_{0}+a_{1}x+a_{2}x^{2}+a_{3}x^{3}+a_{4}x^{4}
\end{equation}
and find that $a_{0}=+1.93354$, $a_{1}=+0.50580$, $a_{2}=-0.01227$,
$a_{3}=-0.02425$ and $a_{4}=0.00850$. At high temperatures where
$x<-1.2$, we use the virial expansion form, 
\begin{equation}
g_{2}\left(x\right)\simeq\frac{1+2b_{2}e^{x}+3b_{3}e^{2x}}{1-2^{-3/2}e^{x}+3^{-3/2}e^{2x}},
\end{equation}
where $b_{2}=3\sqrt{2}/4$ and $b_{3}\simeq0.29095297$ are the second
and third virial coefficients of a unitary Fermi gas \cite{Liu2009,Liu2013},
respectively. To make $g(x)$ smooth across the whole normal state,
we connect $g_{1}(x)$ and $g_{2}(x)$ by using the expression, 
\begin{equation}
g_{n}\left(x\right)=\frac{g_{1}\left(x\right)}{1+e^{-4\left(x+1.2\right)}}+\frac{g_{2}\left(x\right)}{1+e^{+4\left(x+1.2\right)}}.
\end{equation}
Recalling that $g_{s}\left(x\right)=g_{n}(x)$ at $x_{c}\simeq2.49$,
we therefore set $x_{c}=2.49451942$. We note that, we do not need
to impose the constraint $g'_{s}(x_{c})=g'_{n}(x_{c})$, due to the
superfluid phase transition.

\section{Matrix elements of $\mathcal{S}(\omega)$}

In this appendix, we present the weighted mass moments and spring
constants in their dimensionless form, 
\begin{equation}
\tilde{M}_{ij}=M_{ij}\frac{\pi^{5/2}\hbar^{3}\omega_{z}}{m^{2}R_{F}^{2}\left(k_{B}T\right)^{2}}
\end{equation}
and 
\begin{equation}
\tilde{K}_{ij}=K_{ij}\frac{\pi^{5/2}\hbar^{3}\omega_{z}}{m^{2}R_{F}^{2}\left(k_{B}T\right)^{2}}\omega_{\perp}^{2},
\end{equation}
respectively, for the purpose of performing numerical calculations.
Accordingly, we solve the matrix $\mathcal{S}(\omega)$ in its dimensionless
form, 
\begin{equation}
\left[\mathcal{\tilde{S}}\left(\tilde{\omega}\right)\right]_{ij}\equiv\left[\begin{array}{cc}
\tilde{M}_{ij}^{(a)}\tilde{\omega}^{2}-\tilde{K}_{ij}^{(a)} & -\tilde{K}_{ij}^{(ae)}\\
-\tilde{K}_{ji}^{(ae)} & \tilde{M}_{ij}^{(e)}\tilde{\omega}^{2}-\tilde{K}_{ij}^{(e)}
\end{array}\right],
\end{equation}
where $\tilde{\omega}\equiv\omega/\omega_{\perp}$ is the reduced
frequency. For given chemical potential $\mu_{0}/k_{B}T$ or temperature
$T/T_{F}$, these dimensionless matrix elements are given by, after
taking $y=\tilde{r}_{\perp}^{2}$, 
\begin{eqnarray}
\tilde{M}_{ij}^{\left(a\right)} & = & \int_{0}^{\infty}dyy^{\frac{i+j+2}{2}}\left[f_{n}\right],\\
\tilde{M}_{ij}^{\left(e\right)} & = & \int_{0}^{\infty}dyy^{\frac{i+j+2}{2}}\left[\left(1-\frac{f_{s}}{f_{n}}\right)f_{s}\right],
\end{eqnarray}
for the weighted mass moments and 
\begin{eqnarray}
\tilde{K}_{ij}^{\left(a\right)} & = & \frac{3}{2}\frac{\left(i+2\right)\left(j+2\right)}{\left(i+j+2\right)}\tilde{M}_{ij}^{\left(a\right)},\\
\tilde{K}_{ij}^{\left(ae\right)} & = & \frac{i\left(i+2\right)}{4}\frac{T}{T_{F}}\int_{0}^{\infty}dyy^{\frac{i+j}{2}}\left[\bar{s}_{2}f_{s}\right],\\
\tilde{K}_{ij}^{\left(e\right)} & = & \frac{T}{2T_{F}}\int_{0}^{\infty}dyy^{\frac{i+j}{2}}\left[\left(i+2\right)\bar{s}_{2}f_{s}-\frac{2T_{F}}{T}\left(\bar{s}_{2}f_{s}\right)^{'}y\right]\nonumber \\
 &  & \times\left[\left(j+2\right)\bar{s}_{2}f_{s}-\frac{2T_{F}}{T}\left(\bar{s}_{2}f_{s}\right)^{'}y\right]\frac{1}{\left[f_{n}\bar{c}_{v2}\right]},
\end{eqnarray}
for the spring constants. In the above expressions, $ $the argument
for the universal scaling functions is $x=\mu_{0}/(k_{B}T)-y/(T/T_{F})$,
i.e., $[f_{n}]$ is the short-hand notation of $f_{n}[\mu_{0}/(k_{B}T)-y/(T/T_{F})]$,
and $\left(\bar{s}_{2}f_{s}\right)^{'}$ stands for the derivative
$d\left[\bar{s}_{2}f_{s}\right](x)/dx.$

\section{Solving $\det[\tilde{S}(\tilde{\omega})]=0$}

To solve the matrix equation 
\begin{equation}
{\cal \tilde{S}}{\bf \left(\tilde{\omega}\right)}\Lambda=0,\label{matrixeq}
\end{equation}
where the vector of displacement fields, $\Lambda=[A_{0},B_{0}...,A_{i},B_{i},...]^{T}$,
we rewrite the matrix ${\cal \tilde{S}}{\bf (}\tilde{\omega})={\bf M}\tilde{\omega}^{2}-{\bf K}$.
Here, ${\bf M}$ and ${\bf K}$ denote collectively the matrix of
the dimensionless weighted mass moments and the spring constants,
respectively. The matrix ${\bf M}$ is positively definite, so that
we decouple it by a product of a lower triangular matrix ${\bf L}$
and its transpose, 
\begin{equation}
{\bf M=L\cdot L}^{T}.
\end{equation}
In terms of this decomposition, the matrix equation, Eq. (\ref{matrixeq}),
becomes 
\begin{equation}
\left[{\bf L}^{-1}\cdot{\bf K}\cdot\left({\bf L}^{-1}\right)^{T}\right]\cdot{\bf L}^{T}{\bf \Lambda=}\tilde{\omega}^{2}{\bf L}^{T}{\bf \Lambda.}
\end{equation}
It is easy to see that the matrix $[{\bf L}^{-1}\cdot{\bf K}\cdot({\bf L}^{-1})^{T}]$
is symmetric and its eigenvalues give rise to the desired solution
of mode frequencies. The displacement field for each eigenvalue can
also be calculated accordingly, with known eigenstate of the matrix
$[{\bf L}^{-1}\cdot{\bf K}\cdot({\bf L}^{-1})^{T}]$. More explicitly,
if we denote an eigenstate by ${\bf X}$, the corresponding displacement
field is given by 
\begin{equation}
{\bf \Lambda=}\left({\bf L}^{T}\right)^{-1}{\bf X}\text{.}
\end{equation}
Once the displacement fields for a mode are found, we may calculate
its density fluctuation and temperature fluctuation.


\begin{thebibliography}{10}
\bibitem{note1}In the time domain, the displacement fields $\mathbf{u}_{s}$
and $\mathbf{u}_{n}$ are defined by, $\mathbf{v}_{s\perp}(\mathbf{r}_{\perp},t)=\partial\mathbf{u}_{s}(\mathbf{r}_{\perp},t)/\partial t$
and $\mathbf{v}_{n\perp}(\mathbf{r}_{\perp},t)=\partial\mathbf{u}_{n}(\mathbf{r}_{\perp},t)/\partial t$.
\end{thebibliography}
\end{document}